\documentclass[floatfix,twocolumn,nofootinbib,superscriptaddress,a4paper,preprintnumbers]{revtex4-1}

\usepackage{amsmath}
\usepackage{amssymb}
\usepackage{graphicx}
\usepackage[mathscr]{eucal}
\usepackage{color}
\usepackage[T1]{fontenc} 
\usepackage{slashed}
\usepackage{xspace}
\usepackage{hyperref}
\usepackage{xspace}

\newcommand{\lsim}{\stackrel{<}{\sim}}
\newcommand{\sfd}[1]{{\bf #1}}
\newcommand{\sfb}[1]{{\bf {\bar #1}}}
\newcommand{\spa}[1]{\ensuremath{\widetilde{#1}}}
\newcommand{\lam}{\lambda}
\newcommand\SARAH{{\tt SARAH}\xspace}
\newcommand\SPheno{{\tt SPheno}\xspace}
\newcommand{\hc}{\text{h.c.}}
\newcommand{\Veff}{V_{\text{eff}}}
\newcommand{\GeV}{\text{GeV}}
\newcommand{\DRbar}{$\overline{\text{DR}}'$}

\newcommand{\AddrBonn}{%
Bethe Center for Theoretical Physics \& Physikalisches Institut der 
Universit\"at Bonn, \\
 53115 Bonn, Germany
}

\newcommand{\AddrCERN}{%
Theory Division, CERN, 1211 Geneva 23, Switzerland
}

\begin{document}

\hfill BONN--TH--2014--15, CERN-PH-TH-2014-226 \vspace{0.2cm}

\title{On the two--loop corrections to the Higgs mass in trilinear $R$--parity violation}

 \author{Herbi K. Dreiner} \email{dreiner@th.physik.uni--bonn.de}
 \affiliation{\AddrBonn}

  \author{Kilian Nickel} \email{nickel@th.physik.uni--bonn.de}
 \affiliation{\AddrBonn}
 
 \author{Florian Staub}\email{florian.staub@cern.ch}
 \affiliation{\AddrCERN}

% \pacs{14.60.Pq, 12.60.Jv, 14.80.Cp}

\begin{abstract}
We study the impact of large trilinear $R$--parity violating couplings on the lightest CP--even 
Higgs boson mass in supersymmetric models. We use the publicly available computer codes
\SARAH and \SPheno to compute the leading two--loop corrections. We use the effective 
potential approach. For not too heavy third generation squarks ($\spa m\lsim1\,\text{TeV}$) and
couplings close to the unitarity bound we find positive corrections up to a few GeV in the Higgs 
mass.
\end{abstract}
\maketitle

\section{Introduction}
On July 4$^{\text{th}}$, 2012 the discovery of the Higgs boson was announced at CERN 
\cite{Aad:2012tfa,Chatrchyan:2012ufa}. It is not yet established whether this is the Standard 
Model (SM) Higgs boson \cite{Bechtle:2014ewa,Englert:2014uua,Belanger:2013xza}.
However, in the SM the Higgs sector suffers from the hierarchy problem \cite{Veltman:1980mj}, 
to which supersymmetry (SUSY) \cite{Nilles:1983ge,Martin:1997ns} is the most obvious 
solution. It predicts a wide range of observables at the Large Hadron Collider (LHC), for which 
the first run has finished; Run II is expected to start in the Spring, 2015.

There is no convincing experimental indication of any physics beyond the Standard Model
(SM) at the LHC\footnote{See for example the talk given by O.~Buchm\"uller at the EPS 2013 conference in 
Stockholm https://indico.cern.ch/event/218030/ session/28/contribution/869/material/slides/.}\!\!. 
This puts pressure on many proposed scenarios for beyond the standard model (BSM) physics, in 
particular also SUSY. The simplest SUSY scenario, the constrained minimal supersymmetric 
Standard Model (CMSSM) \cite{Nilles:1983ge}, is now excluded \cite{Bechtle:2014yna}, see also 
\cite{Bechtle:2013mda,Craig:2013cxa,Bechtle:2012zk,Buchmueller:2012hv}. However, the MSSM 
extended for example by $R$--parity violation ($R$pV) operators 
\cite{Hall:1983id,Barbier:2004ez,Dreiner:1997uz,Bhattacharyya:1997vv,Allanach:2003eb} can 
significantly weaken the collider mass limits 
\cite{Allanach:2012vj,Asano:2012gj,Franceschini:2012za,Evans:2012bf} and provide an even richer 
phenomenology than the MSSM 
\cite{Dreiner:1991pe,Allanach:2006st,Allanach:2006st,Dreiner:2011wm,Dreiner:2012np,Dreiner:2012wm}. 

Within SUSY the mass of the Higgs boson is restricted at tree--level to be less than the mass of the 
$Z^0$--boson.  However realised, the quantum corrections to the mass can be large 
\cite{Haber:1990aw,Ellis:1990nz}. The observed mass of the Higgs boson, $m_h^{\text{exp}}\approx 
125.7$~GeV \cite{Aad:2014aba,Khachatryan:2014ira,Agashe:2014kda}, is well within the previous 
predicted allowed range for SUSY models \cite{Carena:2000dp}. Such large corrections however 
typically require very large mixing in the stop sector and/or a very heavy stop squark. This in turn is 
disfavoured by fine--tuning arguments \cite{Dimopoulos:1995mi,Birkedal:2004xi}.

When extending the MSSM these conclusions can be modified, \textit{e.g.} in the 
NMSSM \cite{Cao:2012fz,King:2012is,Gunion:2012zd}. Here we consider the Higgs 
mass in supersymmetric models with $R$pV. The additional operators contribute to the 
Higgs mass at the two--loop level\footnote{See also the two--loop $R$pV renormalization 
group equations, which modify the running of the Higgs mass \cite{Allanach:1999mh}.}\!\!. 
This effect is expected to be large especially when involving 
third generation squarks. We study the impact of large $\sfd L\sfd
Q\sfb D$ and  $\sfb U\sfb D\sfb D$ operators involving stops and sbottoms on the lightest 
CP--even Higgs boson mass. (The effects of $\sfd L\sfd L\sfb E$ are here completely 
negligible.) For this purpose we 
calculate two--loop Higgs masses in models beyond the MSSM, but with MSSM precision, 
with the public computer tools \SARAH 
\cite{Staub:2008uz,Staub:2009bi,Staub:2010jh,Staub:2012pb,Staub:2013tta}  and \SPheno 
\cite{Porod:2003um,Porod:2011nf}, as recently presented in \cite{Goodsell:2014bna}.

This letter is organized as follows: we present in the next section our conventions for the models we 
consider, before we give details about the two--loop calculation in sec.~\ref{sec:twoloopcalc}. The 
numerical results are presented in sec.~\ref{sec:results}, before we conclude in sec.~\ref{sec:conclusion}.

\section{The MSSM extended by trilinear $R$--parity violation}
\label{sec:model}
$R$--parity  is a discrete multiplicative $Z_2$ symmetry of the MSSM, defined as 
\cite{Farrar:1978xj,Hall:1983id,Barbier:2004ez,Dreiner:1997uz,Allanach:2003eb}
\begin{equation}
  \label{eq:RParity}
  R_P = (-1)^{3(B-L)+2s} \,,
\end{equation}
where $s$ is the spin of the field and $B$, $L$ are its baryon respectively lepton number. We consider 
the $R$-parity conserving superpotential of the MSSM
\begin{eqnarray}
 W_R&=& Y^{i j}_{e}\,\sfd L_i \sfb E_j \sfd H_d 
   +Y^{i j}_{d}\, \sfd Q_i \sfb D_j \sfd H_d 
   \nonumber \\ &&+ 
   Y^{i j}_{u} \sfd Q_i \sfb U_j \sfd H_u +
 \mu\, \sfd H_u \sfd H_d \, ,
 \label{eq:superpot}
\end{eqnarray}
and extend it by trilinear $R$pV operators 
\cite{Weinberg:1981wj,Sakai:1981pk}
\begin{equation}
  \label{eq:superpotRpV}
 W_{\slashed R} = \frac 12\lam_{ijk} \sfd L_i \sfd L_j \sfb E_k  + \lam_{ijk}^{\prime} \sfd L_i 
 \sfd Q_j \sfb D_k  + \frac 12\lam_{ijk}^{\prime\prime} \sfb U_i \sfb D_j \sfb D_k \,.
\end{equation}
We assume the bi--linear term has been rotated away \cite{Dreiner:2003hw}. Here $i,j,k=1,2,3$ 
are generation indices, while $SU(3)$ colour and $SU(2)$ isospin indices are suppressed. Above
$\sfd L_i,\, \sfb E_j,\, \sfd Q_i,\, \sfb U_i,$ $\sfb D_i$, $\sfd H_d$, 
$\sfd H_u$ denote the left chiral superfields of the MSSM in the standard notation 
\cite{Allanach:2003eb}.  We thus have for the total superpotential
\begin{equation}
W_{\mathrm{tot}}=W_R + W_{\slashed R}\,.
\label{superpot}
\end{equation}
In the following we consider only the presence of one $R$pV operator at a time, 
including the anti-symmetric counter part, if it exists. This ensures the stability of the proton and 
avoids many constraints from flavour changing neutral currents and lepton flavour violation 
\cite{Barger:1989rk,Agashe:1995qm,Dreiner:2003hw}.

The corresponding standard soft supersymmetry breaking terms for the scalar fields $\spa L,
\spa E,\spa Q, \spa U,\spa D, H_d$, $H_u$ and the gauginos $\spa{B},\spa{W},\spa{g}$ read
\begin{eqnarray}
 -\mathscr{L}_{\text{SB},R} &=& m_{H_u}^2 |H_u|^2 + m_{H_d}^2
|H_d|^2+ \sum_\ell \spa{\phi}_\ell^\dagger m_{\spa{\phi}_\ell}^2 \spa{\phi}_\ell \nonumber \\ 
& +& \hspace{-0.2cm}\frac{1}{2}\left(M_1 \, \spa{B}
\spa{B} + M_2 \, \spa{W}_a \spa{W}^a + M_3 \, \spa{g}_\alpha
\spa{g}^\alpha + \hc \right) \nonumber \\ 
&+& (\spa{Q}
T_u\spa{U}^\dagger H_u +  \spa{Q} T_d \spa{D}^\dagger H_d + 
\spa{L} T_e \spa{E}^\dagger H_d \nonumber \\[2mm]
&+& B_\mu H_u H_d + \hc) 
\label{softterms} \\
-\mathscr{L}_{\text{SB},\slashed R} &=& \frac 12 T_{\lam,{ijk}} \spa L_i
 \spa L_j \spa E_k+ T_{\lam,{ijk}}^{\prime} \spa L_i
 \spa Q_j \spa D_k \nonumber \\
 && + \frac 12 T_{\lam,{ijk}}^{\prime\prime} \spa U_i
 \spa D_j \spa D_k + \hc \,. \label{eq:rpv-softterms}
\end{eqnarray}
with $\tilde\phi_\ell \in \{\spa{Q},\spa{D},\spa{U},\spa{E},\spa{L}\}$. The gaugino fields are 
two component fermions \cite{Dreiner:2008tw}. We have suppressed all generation indices
in Eq.~(\ref{softterms}). The $m_{\tilde \phi}^2$ are 3$\times$3 matrices and denote the 
squared soft masses of the scalar components $\tilde \phi$ of the corresponding chiral superfields 
$\Phi$. The $T_{u,d,e}$ are  3$\times$3 matrices of mass--dimension one. They can be written in terms of the standard $A$--terms 
\cite{Nilles:1982dy}, if no flavour violation is assumed, $T_{ii}^f= A_{i}^f Y_f^{ii}$, $f=e,u,d$, $i=1,2,3$, and
no summation over repeated indices. Similarly, for the baryon number violating term
we have $T^{''}_{\lam,ijk}=A^{''}_{ijk}\lam^{''}_{ijk}$.

\section{Two--loop corrections from $R$--parity violating operators}
\label{sec:twoloopcalc}
In the presence of trilinear $R$pV there are new contributions to the Higgs mass at the
two--loop level. We use the public codes \SARAH and \SPheno to compute them. 
These codes perform an effective potential calculation based on the generic results  in 
Ref.~\cite{Martin:2001vx} in the \DRbar scheme. The precision of this
calculation using \SARAH and \SPheno is the same for models beyond the MSSM as in 
many public computer tools for the MSSM, by using the results of 
Refs.~\cite{Brignole:2001jy,Degrassi:2001yf,Brignole:2002bz,Dedes:2002dy,Dedes:2003km}. For 
more general information about the calculation of two--loop Higgs masses in extensions of the MSSM 
with \SARAH and \SPheno we refer to Ref.~\cite{Goodsell:2014bna}.

\begin{figure}[hbt]
\includegraphics[width=\linewidth]{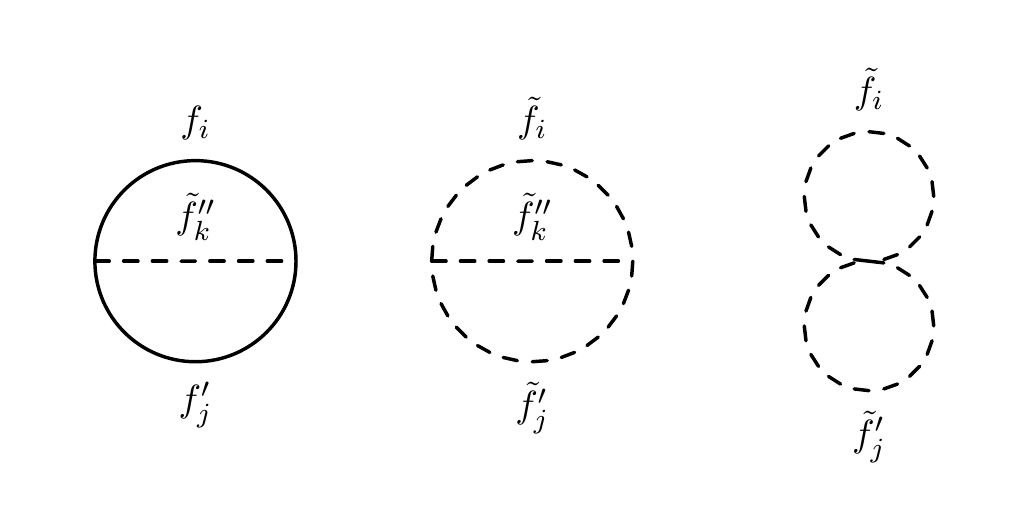} 
 \caption{Two--loop corrections to the effective potential involving trilinear $R$pV
 couplings. $f$ are SM fermions and $\tilde{f}$ are SUSY sfermions. The graph on the 
 left involves superpotential couplings, Eq.~(\ref{eq:superpotRpV}), the middle graph involves
 soft supersymmetry breaking terms, Eq.~(\ref{eq:rpv-softterms}), and the graph on the right,
$R$pV terms in the $F$--term scalar potential. }
\label{fig:Diagrams}
\end{figure}

The corrections to the effective potential at the two--loop level involving trilinear $R$pV couplings 
come from the diagrams shown in Fig.~\ref{fig:Diagrams}.
From these, the tadpole contributions and self--energies are calculated by taking the 
first and second derivative of the two--loop effective potential $\Veff^{(2)}$
\begin{eqnarray}
\label{eq:dti2}
\delta t^{(2)}_i &=& \frac{\partial \Veff^{(2)}}{\partial v_i} \,, \\
\label{eq:pi2}
\Pi_{h_i h_j}^{(2)}(0) &=& \frac{\partial^2 \Veff^{(2)}}{\partial v_i \partial v_j} \,,
\end{eqnarray}
with $i=u,d$. Here, $h_i$ are the real parts of the neutral Higgs 
scalar fields, $H_{u,d}^0$, with $H_i^0=(v_i+h_i+i\sigma_i)/\sqrt 2$.
There are two possibilites to take the derivatives: either calculate numerically the 
derivative of the entire potential as done in Ref.~\cite{Martin:2002wn} for the MSSM, or 
take analytically the derivative of the potential with respect to the masses and 
numerically the derivative of the masses and couplings with respect to the VEVs 
(semi--analytical approach). The combination \SARAH/\SPheno has implemented both 
methods and we check their numerical agreement. Throughout we neglect the 
possibility of sneutrino vacuum expectation values for the $\sfd L\sfd Q
\sfb D$ operators. These effects are very small since the bounds on neutrino masses 
restrict the sneutrino VEVs to be of order 10 MeV or smaller \cite{Allanach:2003eb}.

We use the results of Eqs.(\ref{eq:dti2}) and (\ref{eq:pi2}) together with the tree--level 
minimization conditions, $T_i$, and the one--loop corrections to find the minimum of the 
effective potential by demanding
\begin{equation}
 T_i + \delta t_i^{(1)} + \delta t_i^{(2)} = 0
\end{equation}
and to calculate the loop corrected Higgs mass matrix squared
\begin{equation}
M^2_h(p^2) = [M_h^{(T)}]^2 - \Pi_{h_i h_j}^{(1)}(p^2) - \Pi_{h_i h_j}^{(2)}(0).
\end{equation}
$[M_h^{(T)}]^2$ is the Higgs mass matrix squared at tree-level at the 
minimum of the effective potential. The two eigenvalues $m_{h_i}^2$ of $M^2_h(p^2=m_{h_i}^2)$,
$i=1,2$, are the pole masses of the corresponding scalar fields. The smaller eigenvalue, $m_{h} 
\equiv m_{h_1}$, is the mass of the SM--like Higgs boson, which we are mainly interested in.

In addition to the two--loop corrections to the Higgs potential due to trilinear $R$pV parameters, 
there are also one--loop corrections to the SM Yukawa couplings due to the trilinear $R$pV 
parameters, see for example \cite{Allanach:1999mh}. In particular there are one--loop $R$pV 
contributions to the up and down quark self--energy matrices: $\Sigma_L^{q}$, 
$\Sigma_R^{q}$, $\Sigma_S^{q}$, $q=u,d$. These self--energies in turn contribute at 
one--loop to the Higgs potential, leading to an overall two--loop effect on the Higgs mass, \textit{i.e.} 
of the same order as we are investigating.  These self-energies enter the calculation of the Yukawas 
couplings as \cite{Pierce:1996zz}
\begin{eqnarray}
\frac{v_q}{\sqrt{2}} Y^q &=&  U_L^T m^{q,\text{pole}} U_R + \Sigma_S^{q} +\Sigma_L^{q,T} \left(\frac{v_q}{\sqrt{2}} Y^q 
\right) \nonumber \\
 &&+    \left(\frac{v_q}{\sqrt{2}} Y^q \right) \Sigma_R^{q} + \dots\,,
\end{eqnarray}
which has to be solved iteratively. The dots stand for 
two--loop corrections important for the top quark, $U_L$, $U_R$ are the matrices which 
diagonalize the Yukawa matrix $Y^q$. $m^{q,\text{pole}}$ is a diagonal matrix with the 
pole masses as entries.

\section{Results}
\label{sec:results}
\begin{figure}[h!]
 \includegraphics[width=0.9\linewidth]{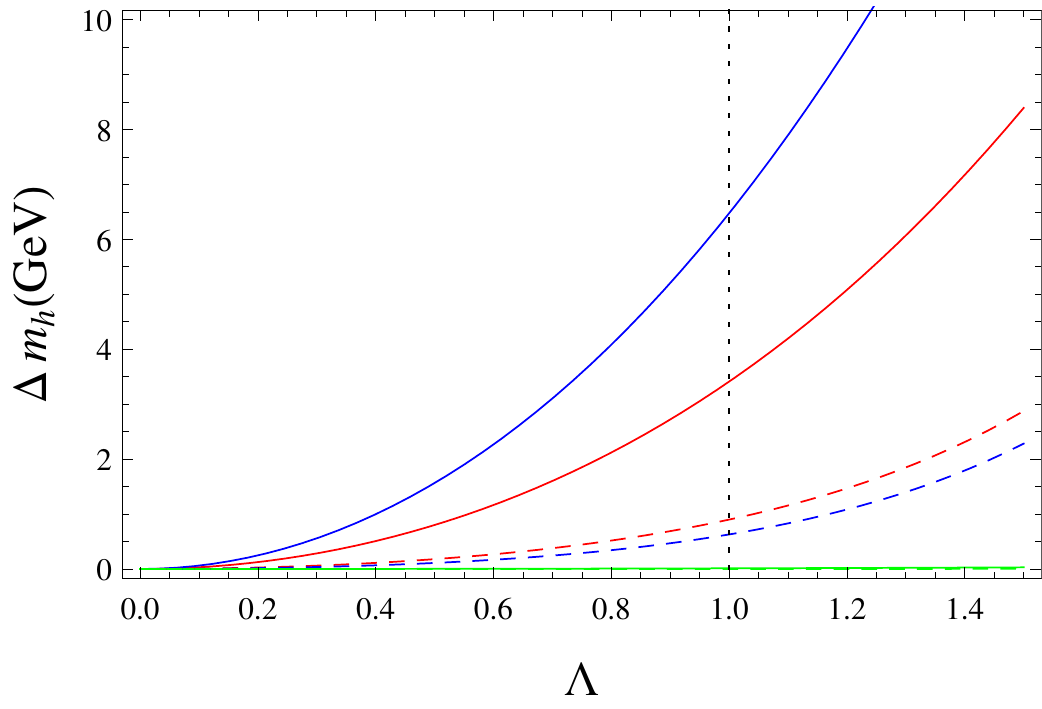} \\[1mm]
 \includegraphics[width=0.9\linewidth]{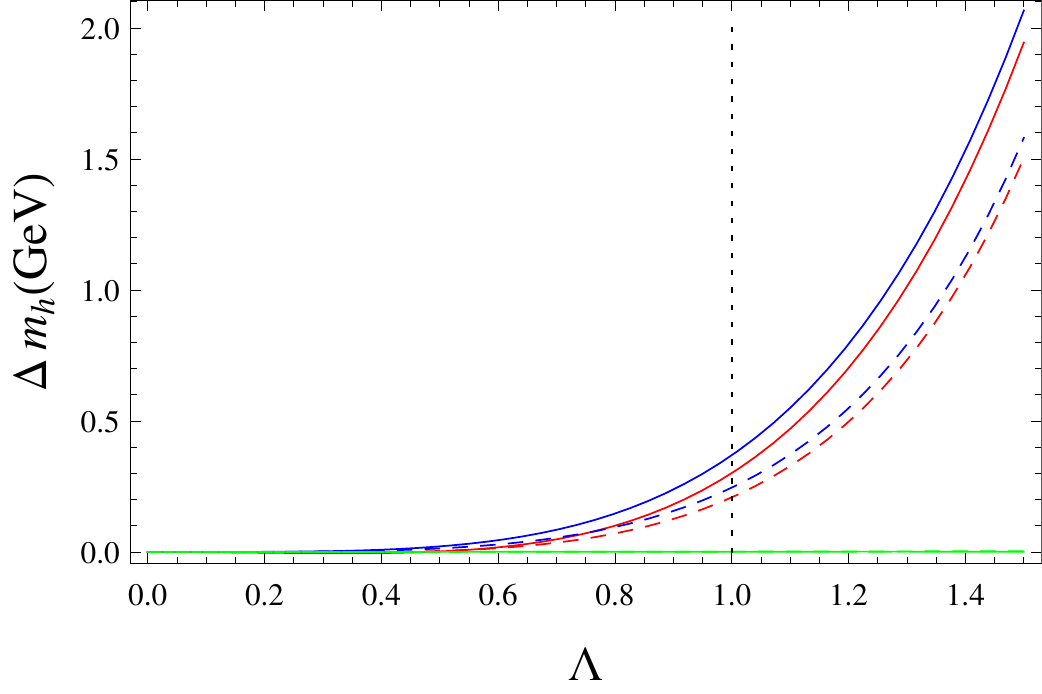} 
 \caption{$\Delta m_h$ for the two mass hierachies given at the beginning of Sect.~\ref{sec:results},
 hierarchy  (i)~[top plot], and (ii) [bottom plot]. The shift is shown as a function of of $\Lambda = 
 \lam^{'}_{ijk}, \lambda^{''}_{ijk}$, with the color code: $\lam^{''}_{313}$ (full red), $\lambda^{''}_{312}$ 
 (full blue), $\lam^{''}_{213}$ (full green), $\lam^{'}_{333}$ (dashed red), $\lam^{'}_{331}$ (dashed blue), 
 $\lam^{'}_{313}$ (dashed green). The dashed, vertical line indicates the perturbativity limit. The two green 
 lines are degenerate in both plots.}
\label{fig:results}
\end{figure}

We now discuss the numerical impact of the $R$pV operators on the Higgs mass at the two--loop level. 
To be specific, we consider the supersymmetric parameter point fixed by $\tan\beta=10$, $M_1 
= M_2 = \frac{1}{2} M_3 = 1$~TeV, $\mu=0.5$~TeV, and $M_A =1$~TeV. All slepton soft masses as well
as all squark soft masses  of the first two generations are set to $1.5~\text{TeV}\!$. For the third 
generation squarks soft masses we distinguish two exemplary mass hierarchies
\begin{itemize}
 \item[(i)] $m_{\tilde{Q},33} = 1.5~\text{TeV}$, \; \;\;\;$m_{\tilde{U},33} = m_{\tilde{D},33} = 0.5~\text{TeV}$\,,
% \item[(ii)] $m_{\tilde{Q},33} = m_{\tilde{U},33} = m_{\tilde{D},33} = 1.5~\text{TeV}$\,,
 \item[(ii)] $m_{\tilde{Q},33} = m_{\tilde{U},33} = m_{\tilde{D},33} = 2.5~\text{TeV}$\,.
\end{itemize}
In (i) the third generation is lighter than the other sfermions, in (ii) it is heavier. The two hierarchies are 
assumed in the two plots shown in Fig.~\ref{fig:results}. We choose the $R$-parity conserving trilinear parameters as $T_t = -2.5$~TeV, resulting in large mixing in the stop sector; all other $R$-parity conserving trilinear parameters vanish. In the $R$pV sector we choose
\begin{equation}
 T_{\Lambda_{ijk}} = A_0 \Lambda_{ijk}\,, \qquad \Lambda=\lam',\lam''\,,
\end{equation}
with $A_0= -2.5$~TeV. The renormalization scale is always set to $Q=\sqrt{m_{\tilde{t}_1} 
m_{\tilde{t}_2}}$, where $m_{\tilde{t}_i}$ are the \DRbar stop masses. For the SM parameters we
use $m_t = 173.1$~GeV, $m_b = 4.18$~GeV, $m_\tau = 1.777$~GeV, and $\alpha_S = 0.1184$.
The impact on the light Higgs mass as a function of the $R$pV trilinear couplings $\Lambda$ is defined as
\begin{equation}
 \Delta m_h {\,\equiv\,} m_h(\Lambda) - m_h(0)\,,
\end{equation}
where the Higgs mass in the $R$-parity conserving case, $m_h(0)$, for the two hierarchies
 is given by
\begin{itemize}
\item[(i)] $m_h(0)= 110.0 \,\GeV\,,$
%\item[(ii)] $m_h(0)=  125.5\,\GeV\,,$
\item[(ii)] $m_h(0)= 124.3 \,\GeV$.
\end{itemize}
Since we just wish to demonstrate an effect, we have not attempted to tune our parameters to get 
the correct Higgs mass in all scenarios. We restrict ourselves to the couplings $\lam^{''}_{313}$, 
$\lam^{''}_{312}$, $\lam^{''}_{213}$, $\lam^{'}_{313}$, $\lam^{'}_{331}$, and  $\lam^{'}_{333}$. 
However through the radiative corrections we dynamically generate further couplings. As 
mentioned, since the operators corresponding to $\lam_{ijk}$ do not couple to squarks, the 
associated corrections to the Higgs mass are negligible. For the green line in the two plots of 
Fig.~\ref{fig:results}, this is also the case, corresponding to squark contributions not involving stops: 
$\lam''_{213},\,\lam'_{313}$. 

In general, we find that for light third generation squarks, hierarchy (i), shown in the top plot in 
Fig.~\ref{fig:results}, there can be large positive contributions of several GeV to the Higgs mass, 
if stops are involved in the $R$pV operator. If the third generation squarks are heavier (hierarchy (ii))  
shown in the bottom plot in Fig.~\ref{fig:results}, the effects are significantly smaller. 

To get large effects, the $R$pV couplings have to be very large. An enhancement of several GeV is only 
found for couplings which are close to or even above the perturbativity limit, which is approximately 1 at 
the weak scale \cite{Brahmachari:1994wd,Allanach:1999ic}\footnote{The authors required perturbativity, 
or lack of a Landau pole, up to the unification scale $M_X\approx 10^{16}\,\text{GeV}$. The bounds are 
given at the weak scale.}\!. In order to avoid the Landau pole the large coupling scenarios must have 
a low cut--off similar to the $\lambda$--SUSY setup \cite{Hall:2011aa}. 

The couplings involving stops are hardly constrained by flavor physics, especially if the non--stop masses 
are in the TeV range~\cite{Dreiner:2013jta}. Furthermore, we have checked that for example for $\lam''_{312}$ if we choose instead $\tan\beta=25$, the resulting shift in the Higgs mass changes by less 
than 5\%. We have to note that very small soft masses together with large trilinear couplings often suffer 
from an unstable electroweak vacuum and have to be considered carefully 
\cite{Camargo-Molina:2013sta,Camargo-Molina:2014pwa,Chamoun:2014eda}. We used the public code 
{\tt Vevacious} \cite{Camargo-Molina:2013qva} to check that hierarchy (i) is meta-stable with a life--time 
longer than the age of the universe.

\begin{figure}[hbt]
\vspace{0.4cm}
  \includegraphics[width=0.9\linewidth]{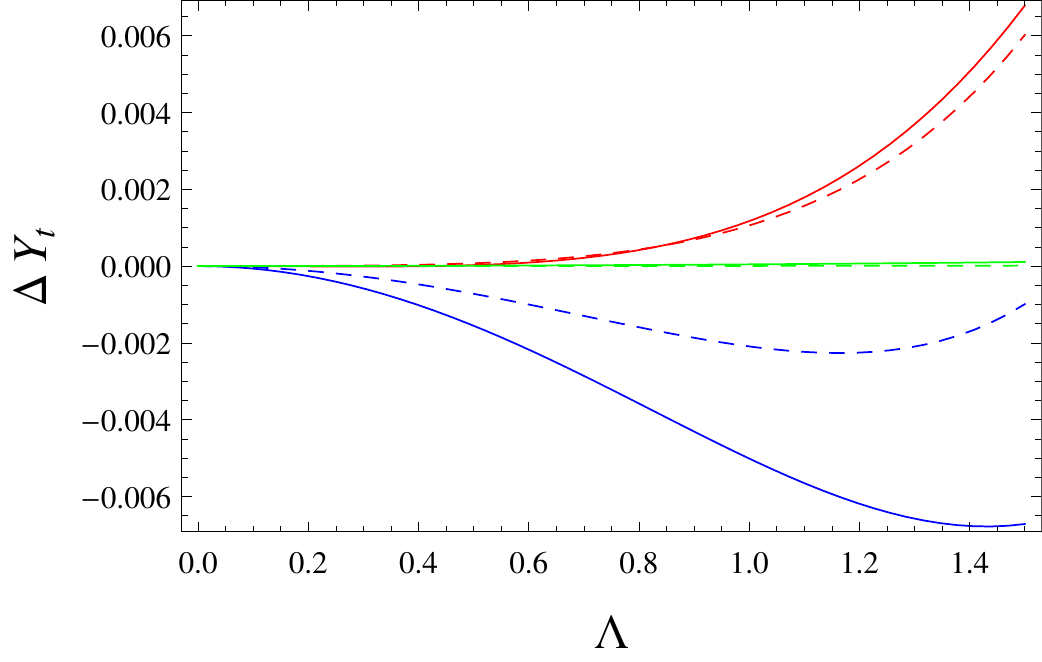} 
 \caption{The change in the top Yukawa coupling, $\Delta Y_t(\Lambda)$,  for the first mass 
 hierarchy given at the beginning of Sect.~\ref{sec:results}. The color code is the same as for 
 Fig.~\ref{fig:results}.}
 \label{fig:DeltaYukawas}
\end{figure}

\begin{figure}[hbt]
 \includegraphics[width=0.9\linewidth]{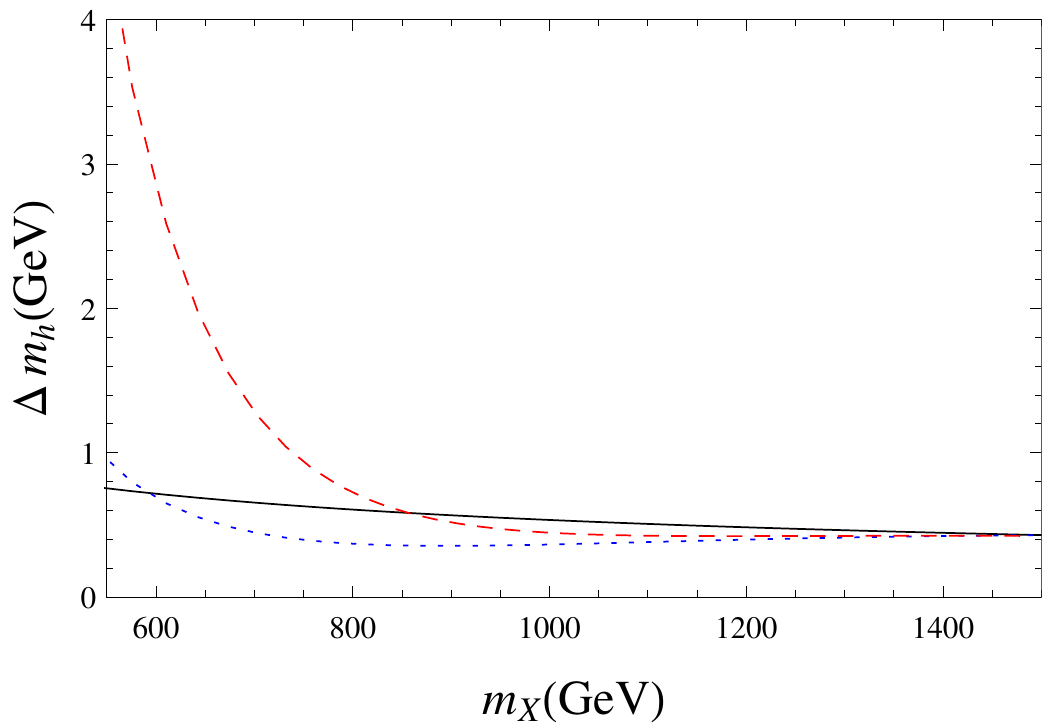} \\[4mm]
 \includegraphics[width=0.9\linewidth]{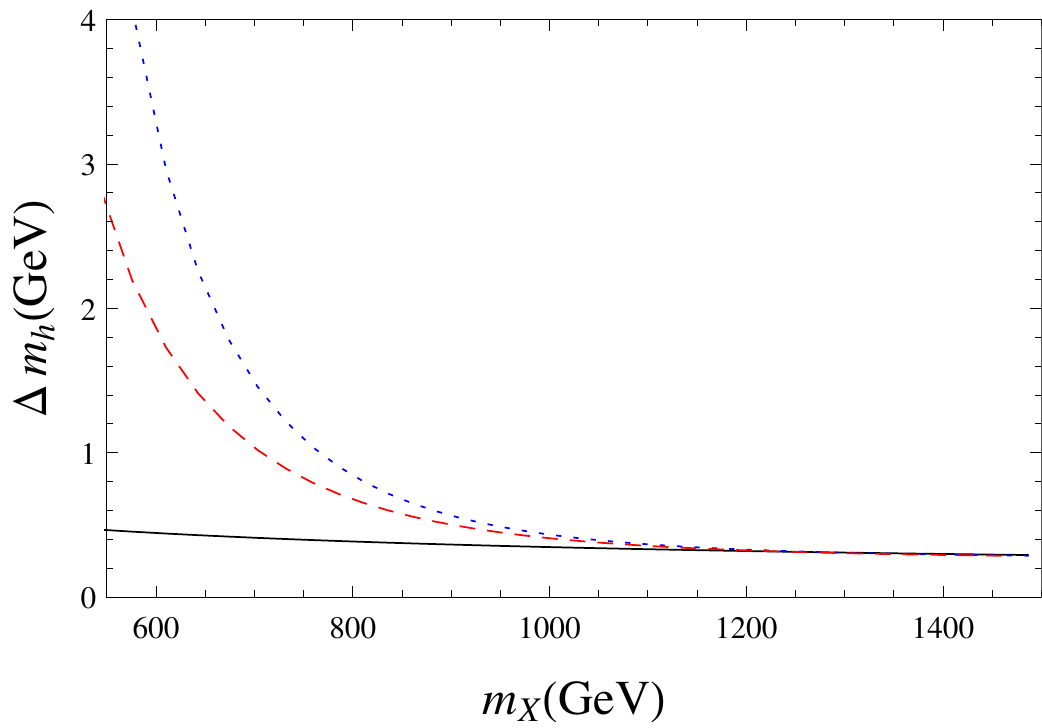} 
 \caption{Two--loop $R$pV contributions to the light Higgs mass as a function of the soft
 squark masses. We set all soft masses to be 1.5~TeV and then vary $m_{\tilde{Q}}$ 
 (blue), $m_{\tilde{U}}$ (red), and $m_{\tilde{D}}$ (black), independently, while keeping the 
 other masses fixed. In the first plot, $\lam^{''}_{313} = 1$ and 
 $T_{\lam^{''}_{313}} =-2.5$~TeV, in the second $\lam^{'}_{333} =1$ and $T_{\lam^{'}_
 {333}} =-2.5$~TeV.}
 \label{fig:mX2dependence}
\end{figure}

We show in Fig.~\ref{fig:DeltaYukawas} the change in the top--Yukawa coupling 
\begin{equation}
\Delta Y_t(\Lambda) \equiv Y_t(\Lambda) - Y_t(0)\,, 
\end{equation}
from including the $R$pV loop corrections to all quarks. Here $Y_t(0) \simeq 0.85$, for $\tan\beta
=10$. The effect is very small. The dependence of the Higgs mass on the mass of the involved 
squarks is depicted in Fig.~\ref{fig:mX2dependence}, where we kept $\lam^{''}_{313} = 1$, 
respectively $\lam^{'}_{333}=1$, fixed and varied $m_{\tilde{Q},33}$,  $m_{\tilde{U},33}$,  and 
$m_{\tilde{D},33}$, separately. The soft masses not being varied are fixed at 1.5 TeV.

The largest corrections appear in the case of light right--handed squarks together with large 
$\sfb U \sfb D \sfb D$ operators. For $\sfd L \sfd Q \sfb D$ operators the strongest dependence 
is on the left--squark soft mass. The  value of $m_{\tilde{D},33}$ plays always a subdominant 
role.

\begin{figure}[hbt]
 \includegraphics[width=0.9\linewidth]{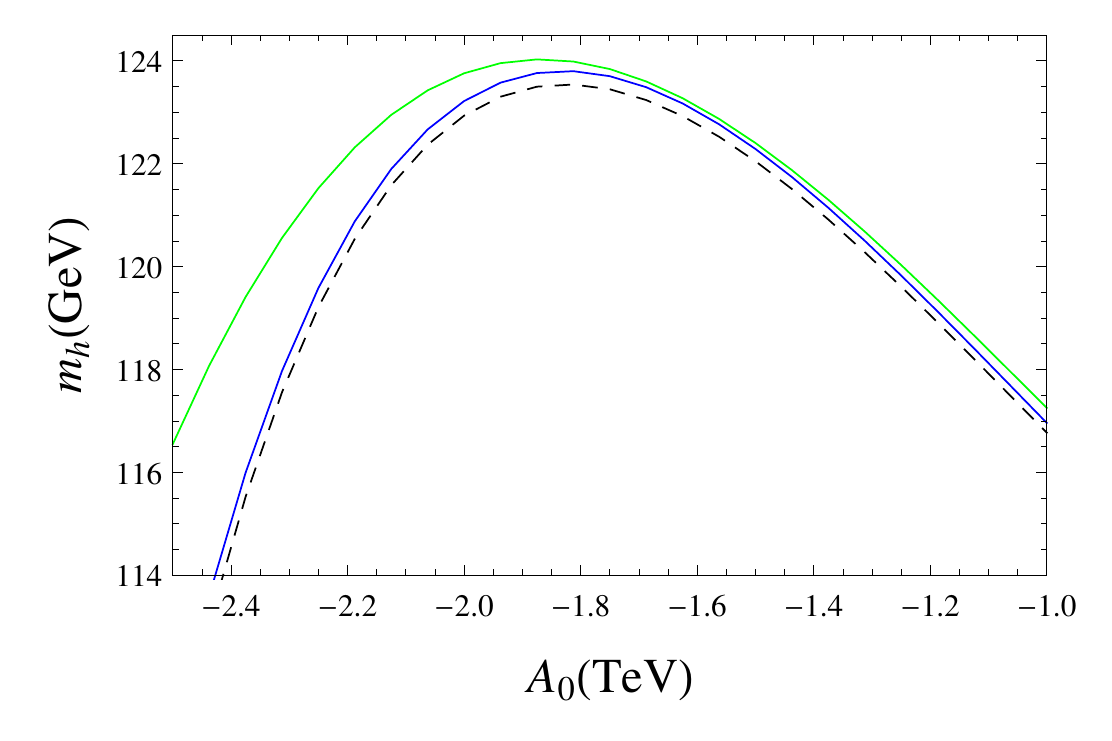} \vspace{-0.5cm}
 \caption{The CP--even Higgs mass $m_h$ as a function of $A_0$. The dashed line is the 
 calculation without $R$pV contributions, while the blue line is for $\lam^{'}_{333} = 1$, $T_{\lam^{'}
 _{333}} = A_0$ and the green one for $\lam^{''}_{313} = 1$, $T_{\lam^{''}_{313}} = A_0$. All sfermion 
 soft masses but $m_{\tilde{U},33}$ are fixed to 1.5~TeV. We set $m_{\tilde{U},33}$ to 0.5~TeV.}
  \label{fig:A0}
\end{figure}

We finally consider the dependence on $A_0$. For this purpose we show in Fig.~\ref{fig:A0}
the light Higgs mass as function of $A_0$ with and without $R$pV operators. Here, we have 
chosen light right--handed stops, $m_{\tilde{U},33}=0.5$~TeV, while all other scalar 
soft masses are set to 1.5 TeV. Once again the $R$pV couplings can easily shift the 
light Higgs mass by a few GeV. In the case of  $\lam^{''}_{313}$ the shift shows a clear 
dependence on $A_0$ while it is rather insensitive to $A_0$ if $\lam^{'}$ couplings are 
considered. That is consistent with our choice of small $m_{\tilde{U},33}$. For small $m_
{\tilde{Q},33}$ the $\lam^{'}$ would show a stronger dependence on $A_0$. 

\section{Conclusion}
\label{sec:conclusion}
We have discussed the impact of large trilinear $R$pV couplings on the light 
CP--even Higgs mass at the two--loop level. We have shown that in particular for light stops 
these corrections can be very important, increasing the Higgs mass by several GeV, 
if the couplings are $\mathcal{O}(1)$. 

\section*{Acknowledgements}
We thank Mark Goodsell for his support in implementing the two--loop corrections in \SARAH and 
many helpful discussions.  We thank Howard Haber for discussions. We are in debt to 
Pietro Slavich who pointed out the relevance of 1--loop $R$pV corrections to the 
SM Yukawas couplings.

\vfill

%\vspace{4cm}

%.

% \bibliographystyle{JHEP}
\bibliography{main}

\end{document}